\documentstyle[aps,epsf,twocolumn]{revtex}
\def\la{\langle} 
\def\ra{\rangle} 
\def\be{\begin{eqnarray}} 
\def\ee{\end{eqnarray}}
\newcommand{\sg}{\sigma}

\begin{document}

\draft

\title{\bf Chiral Disorder in QCD}

\author{ {\bf Romuald A. Janik}$^1$, {\bf Maciej A.  Nowak}$^{1}$ ,
{\bf G\'{a}bor Papp}$^{2}$ and {\bf Ismail Zahed}$^3$}

\address{$^1$ Department of Physics, Jagellonian University, 30-059
Krakow, Poland.\\ 
$^2$ITP, Univ. of  Heidelberg, Philosophenweg 19, D-69120 Germany \&\\ 
Institute for
Theoretical Physics, E\"{o}tv\"{o}s University, Budapest, Hungary.\\
$^3$Department of Physics, SUNY, Stony Brook, New York 11794, USA.}

\maketitle

\begin{abstract}

Using the Gell-Mann-Oakes-Renner (GOR) relation and
semi-classical arguments, we show that the bulk quark spectrum in QCD 
exhibits a variety of regimes including the ergodic one described by random 
matrix theory. We analyze the quark spectral form-factor in the diffusive and 
ballistic regime. We suggest that a class of chiral transitions in QCD is
possibly of the metal-insulator type, with a universal 
spectral statistics at the mobility edge.

\end{abstract}

{\bf 1.}
The chiral properties of QCD in a small Euclidean volume are constrained
by the way chiral symmetry is spontaneously broken in the infinite volume
limit~\cite{GASSER}. This observation carries to the quark spectrum averaged
over gauge configurations of fixed winding number, in the form of microscopic 
sum rules~\cite{SMILGA}. The sum rules can be rederived using chiral random 
matrix models~\cite{SHURYAK}. The microscopic sum rules reflect on the 
universal properties of the eigenvalue distribution and correlations a quantum 
leap apart~\cite{VERZA}.

The interesting results achieved by random matrix models in describing
the microscopic regime of chiral QCD, prompt us to ask whether we could 
think about other generic or universal aspects of the quark spectrum,
aside from that of random matrix theory. In this letter, we would like to 
provide some new insights to the microscopic regime using semi-classical 
arguments. We will use these new insights to analyze the quark spectral 
form-factor in the diffusive and ballistic regimes (non-universal).
We then argue that a number of chiral transitions in QCD are possibly of the 
metal-insulator type, and suggest that at the edge of the metal-insulator
transition a universal spectral statistics is expected.

\vskip .25cm

{\bf 2.}
To be able to organize the various energy scales of interacting
quarks in a finite Euclidean volume $V$, we call upon the two-flavor
GOR relation~\cite{GOR}, 
\be
F^2  =-
\frac {m}{m_{\pi}^2} \,\,  \la \overline{q} q\ra  = 
\frac {m}{m_{\pi}^2} \,\, \pi \rho (0)  \,,
\label{dis1}
\ee
where $F$ is the pion decay constant, $m_{\pi}$ the pion mass and 
$\la \overline{q}q\ra=-\Sigma$ is the quark condensate. The last identity
in (\ref{dis1}) follows from the Banks-Casher relation~\cite{BANKS}, with 
$\rho(0)$ the quark density of eigenvalues per unit four volume $V=L^4$
at zero virtuality ($\lambda\approx 0$).
Recall that in a metal, the Kubo formula for the dc-conductivity 
$\sigma$ in terms of the diffusion constant $D$ and the density of states
$\rho_F$ at the Fermi level is (in units of $e^2/\hbar$) :
$\sigma = D \rho_F$ \,. By analogy, we see that in QCD, $D\approx F^2/\Sigma$
plays the role of the diffusion constant while
the Fermi level corresponds to the zero virtuality point.
This physical analogy will be confirmed below by a direct calculation.

The appearance of a diffusion constant $D$ allows us
to organize the various stages of the disordered phase.
For a quark with proper time $t\sim 1/\lambda$, 
where $\lambda$ is its virtuality, the relevant time scales are:
the Heisenberg time, $t_H=1/\Delta$ with $\Delta=1/\rho\,V$ the typical
quantum spacing; the ergodic time, $\tau_{\rm erg} =1/E_c= L^2/D$,
with $E_c$ the Thouless energy~\cite{THOULESS}; 
and the diffusion time $\tau_d=1/2m_Q$
with $m_Q$ the constituent quark mass. For fixed $V$, the Ohmic
conductance is $\sigma_L=E_c/\Delta\approx F^2 L^2$.
The relation $D=v_F^2\tau_d/4$
suggests that the virtual velocity of  diffusive quarks at
$\lambda\approx 0$ is about $v_F \approx F(m_Q/\Sigma)^{1/2}\approx 1$.
Diffusive quarks are the analogue of constituent quarks. 

We will refer to the quarks in the regime $t>t_H $ as quantum, those in
the regime $\tau_{\rm erg}<t<t_H$ as ergodic, those in the
regime $\tau_d <t<\tau_{\rm erg}$ as diffusive, and finally those
in the regime $t<\tau_d$ as ballistic. We note that the 
time scales are ordered as $t_H \gg \tau_{\rm erg} \gg \tau_d$ or 
$V \gg \sqrt{V} \gg 1$ in units where $F=\Sigma=m_Q=1$. The virtual quark 
spectrum in QCD bears much resemblance to the real electronic spectrum
in disordered metallic grains~\cite{LEE}.

\vskip .25cm
{\bf 3.}
An important tool for studying the quark level correlations in the
QCD vacuum is the averaged two-level correlation function~\cite{EFETOV}
\be
R_{\Lambda,E}(s) \equiv R(s) = \frac 1{\nu_{\Lambda}^2}\la \nu 
(\lambda_+) \nu (\lambda_- ) \ra -1
\label{disc3}
\ee
measuring correlation between two eigenvalues $\lambda_{\pm}$ in the 
virtuality window around $\Lambda$ with width $E$, being apart from 
each other by  $\Delta s$.  $\nu_{\Lambda}=1/\Delta V$ is the
mean density in the band $E$. 
The averaging in (\ref{disc3}) is over the gauge configurations 
$A$ with the QCD measure. The quark eigenvalues in a fixed gauge 
configuration $A$ are denoted by 
 $i\nabla \!\!\!\!/\,[A] q_n=\lambda_n [A] q_n$,
and the unaveraged density of states is
\be
\nu (\lambda ) = \frac 1V \sum_k \delta (\lambda -\lambda_k[A ] )\,.
\ee
In particular $\rho (0) = \la \nu (0) \ra$. 
Eq.~(\ref{disc3}) is related to the two-point
connected correlation function discussed in \cite{VERZA,NOWAK} for constant
$\nu_{\Lambda}$. It satisfies the sum rule
$\int\!ds R(s) ds =0$ and its Fourier transform is the 
spectral form-factor
\be
K(t) = \int_{-\infty}^{+\infty}
\frac {d\lambda}{2\pi} \, R(\lambda/\Delta) \, e^{-i\lambda t}.
\label{disc4}
\ee

Following the general semi-classical
arguments by Gutzwiller~\cite{GUTZWILLER},
Berry~\cite{BERRY} and others~\cite{OTHERS}, we suggest that for QCD
\be
K( t) \approx \frac{2 |t|\Delta^2 }{(2\pi)^2\beta} \, p(t)\,.
\label{disc5}
\ee
where $p(t)$ is the quark return probability for fixed proper time $t$.
In QCD with $N_c\geq 3$ (number of colors) the quarks are in 
the (complex) fundamental representation. Their evolution operator along the 
proper time does not enjoy time-reversal invariance in the presence 
of chromo-magnetic interactions. In the semi-classical approximation leading
to (\ref{disc5}) this corresponds to $\beta=2$ (only the diagonal terms in
the closed orbit indices are retained).
In QCD with $N_c=2$ the quark orbits are paired by the conjugation operator 
$C\tau_2 K$ \cite{SMILGA}, where $C$ is charge conjugation and $K$ complex 
conjugation, hence $\beta=1$ (diagonal and off-diagonal terms within the pair
retained). In QCD with $N_f$ adjoint and real Majorana quarks, the orbits are 
paired by $CK$ \cite{SMILGA}.
If only one pair of orbits is retained (square root of the theory) then 
$\Sigma=\pi\rho(0)/2$ and $\beta=4$. A similar observation applies to 
Kogut-Susskind quarks on the lattice.

The normalized return probability $p(t)$ of a quark from $x(0)$ back
to $x(t)$ for fixed proper time $t$ in a four volume $V$, is
\be
p(t) = \frac {V^2}N \lim_{y\to x}{}&&
\int \frac {d\lambda_1d\lambda_2}{(2\pi)^2} 
\,e^{-i(\lambda_1-\lambda_2) |t|}\nonumber\\&&
\la {\rm Tr}\left( S(x,y;z_1) S^{\dagger} (x,y; z_2)\right)\ra
\label{des2}
\ee
with $z_{1,2}=m-i\lambda_{1,2}$, and
\be
S(x,y; z) = \la x| \frac 1{i\nabla \!\!\!\!/\, + iz} |y\ra
\label{des3}
\ee
where $N$ is the number of quark states in $V$. 
A proper regularization of the short paths is assumed.
Setting $\lambda_{1,2}=\Lambda\pm \lambda/2$ with $\Lambda\approx 0$
(near the zero virtuality point)~\cite{NOTE1},
then $z_1 =z=m-i\lambda/2$ and 
$z_2=z^*$. In this case, the quark propagator satisfies
\be
S^{\dagger} (x,y; z^*) =-\gamma_5 S(y, x; z)\gamma_5
\label{des4}
\ee
and (\ref{des2}) is seen to relate to the pion correlation
function
\be
\,{\bf 1}^{ab}\, {\bf C}_{\pi} (x,y;z) = 
\la {\rm Tr}\left(
S(x, y;z) i\gamma_5\tau^a S(y,x;z) i\gamma_5\tau^b\right)\ra
\label{des5}
\ee
For $z=m$, pion-pole dominance (long paths) yields
\be
{\bf C}_{\pi} (x,y;m) \approx \frac 1V \sum_Q e^{iQ\cdot (x-y)} 
\frac {|\la \overline{q} q\ra |^2}{F^2} \frac 1{Q^2+m_{\pi}^2}
\label{des06}
\ee
where the sum is over the pion momenta $Q_{\mu}=n_{\mu}2\pi/L$ 
in a symmetric box $V=L^4$. Using the GOR relation (\ref{dis1}) 
and the analytical continuation $m\rightarrow m-i\lambda/2$, we find
\be
{\bf C}_{\pi} (x,y;z) \approx \frac 1V \sum_Q e^{iQ\cdot (x-y)} 
\frac {2|\la \overline{q} q\ra |}{-i\lambda + 2m + DQ^2}
\label{des6}
\ee
with the diffusion constant $D=2F^2/\Sigma$.
Inserting (\ref{des6}) into (\ref{des2}), and noting that
$\int d\Lambda \sim E$ with $E/\Delta=N$ and $\rho=1/\Delta V$,
with $\Sigma=|\la \overline{q} q\ra|=\pi \rho$, we conclude after
a contour integration that
\be
p(t) = 
e^{-2m|t|}\sum_Q e^{-DQ^2 |t|}
\label{des7}
\ee
The prefactor $e^{-2m|t|}$ is the expected damping for zero mode quarks.
Heavy quarks with bare masses $2m>D(2\pi/L)^2$ (symmetric box) 
do not diffuse. 

The present arguments will allow us to analyze the spectral rigidity
\cite{EFETOV,MEHTA}
\be
\Sigma_2 (N) =\int_{-N}^{N}ds\,\, (N-|s|) R(s)\,\,.
\label{def}
\ee
This is the variance in the 
number of virtual quark levels in an energy band $E=N\Delta$, averaged
over the QCD gauge configurations. Its dependence on $N$ or $E$ changes 
qualitatively with the virtuality band considered as we now discuss.

\vskip 0.25cm
{\bf 4.}
In the ergodic regime $\tau_{\rm erg} < t< t_H$ 
the result (\ref{des7}) is dominated
by the constant  mode $n=0$. Hence the return probability is
$p(t)=e^{-2m|t|}$, and
\be
 R(s) = \frac 1{\beta\pi^2} \frac{\alpha^2-s^2}{(\alpha^2+s^2)^2}
\label{disc66}
\ee
with $\alpha=2m/\Delta$. The corresponding spectral rigidity is
\be
\Sigma_2(N) = \frac 1{\beta\pi^2} \,\,{\rm ln} 
\left( \frac {N^2+\alpha^2}{\alpha^2}\right)
\label{disc77}
\ee
for $N\gg 1$. For quark masses in the range 
$\alpha\sim 1$ or $0<m\Sigma V\leq 1$ ($V\sim N$) \cite{GASSER,SMILGA},
the spectral rigidity is universal, $\Sigma_2(N)=2{\rm ln}N/(\beta\pi)^2$.
This is in agreement with the result of standard~\cite{MEHTA} and
chiral~\cite{VERZA,VERBA} random matrix theory.
For quark masses $m\sim 1$ in units where 
$F=\Sigma=1$, the spectral rigidity (\ref{disc77})
deviates from random matrix theory. 
In QCD this mass range is already sensitive to strangeness.

The result is also valid for quenched QCD as (\ref{des06}-\ref{des6})
and the GOR relation hold in this case as well. For $\alpha\sim 1$,
the asymptotic of (\ref{disc66}) is enough~\cite{MEHTA,JALABERT}
to show that the tail
of the bulk level spacing distribution follows that of a Wigner surmise 
\cite{MEHTA}. The arguments in~\cite{JALABERT} can be easily extended 
to general $\alpha$. 

The present arguments only yield the perturbative part (\ref{disc66})
of the two-level correlation function (\ref{disc3}). The oscillatory
part well-known from random matrix theory~\cite{MEHTA}, requires a 
reassessment of (\ref{disc5}) including the
short paths which we have not done in this work. In this context
an interesting observation was made recently by Agam, Altshuler and 
Andreev \cite{AGAM}, relating not only the perturbative part of
(\ref{disc3}) but also the oscillatory part to the spectrum of
the diffusion operator (in our case (\ref{des6})), 
and more generally to the Perron-Frobenius 
operator. A similar observation should hold in the
context of QCD, where the short paths are just controlled by QCD
perturbation theory, thereby establishing 
the chaotic character of the quark spectrum in QCD from first principles.

Finally, we note that the relation (\ref{disc5}) in combination 
with (\ref{des7}) allows us to write 
\be
R(s) =-\frac {\Delta^2}{\beta\pi^2}
{\rm Re}\sum_Q
\frac 1{(s\Delta +iDQ^2 +i2m )^2}
\label{var1}
\ee
in agreement with the perturbative
result derived by Altshuler and Shklovskii
\cite{ALTSHULLER} for disordered electrons in metallic grains. 
In our case, the squared denominator in (\ref{var1}) follows
from the exchange of 2-pions in the double ring diagram 
corresponding to the density-density correlation function (\ref{disc3}). 
This diagram is forced by G-parity and is dominant at large $t$. 
So QCD with $N_c\geq 3$ corresponds to the exchange of 
diffusons~\cite{ALTSHULLER}, hence $\beta=2$.
QCD with $N_c=2$ and $N_f\geq 2$ allows for massless pions and baryons. 
In this case, the baryons are the analogue of the cooperons~\cite{ALTSHULLER},
hence $\beta=1$.

\vskip .25cm
{\bf 6.}
In the diffusive
regime $\tau_d<t<\tau_{\rm erg}$ the result (\ref{des7}) receives 
contribution from all modes in the box $V$. The outcome
is still a classically diffusive motion of the quark in d=4, with
a return probability
\be
p(t) =e^{-2m|t|}\frac  V{(4\pi D t)^2}\,.
\label{diff2}
\ee
Setting $1/m>\tau_{\rm erg}$ that is $\Sigma/V<m<D/\sqrt{V}$ and dialing
the virtuality $\lambda$ such that $1/\tau_{\rm erg} <\lambda<1/\tau_d$
will allow us to probe the diffusive part of the quark spectrum in QCD.
In particular, the spectral form-factor
(\ref{disc5}) can be readily found.
Its Fourier transform is seen to diverge for short times. Setting a short
time cutoff at $\tau_d$ (the diffusion time), we obtain
\be
R(s) = -\frac{V \Delta^2}{(2\pi)^4 D^2 \beta}\left( {\bf C} +
{\rm ln}(\Delta\tau_d\, |s|)\right)\,.
\label{diff4}
\ee
where ${\bf C}=0.577$ is Euler's constant.
We note that $\Delta\tau_d=\tau_d/t_H \gg 1$.  
In the diffusive regime the spectral rigidity is
\be
\Sigma_2 ( N)\!=\!
\frac{V \Sigma^2}{4\beta(2\pi)^4 F^4} 
	\,\left( 3-{\bf C}-{\rm ln} (N\Delta\tau_d)\right)\,(N\Delta)^2
\label{diff5}
\ee
where $N =E/\Delta \gg 1$ is the mean number of states in the energy
band $E$. In the diffusive regime $E\tau_d <1$ so the spectral rigidity
is always positive. However, it {\it decreases} with increasing $N$ or
$E=N\Delta$. Although not universal,
this result should be specific to the QCD quark spectrum in the quoted
regime, and through the $1/F^4$ behavior amenable to 
chiral power counting \cite{LEUTWYLER}.

\vskip 0.25cm
{\bf 7.}
In the ballistic regime
$t<\tau_d$ the classical trajectories of the quarks are shorter
than the typical mean free path 
$l=v_F\tau_d =v_F/2m_Q$, which is about the 
constituent quark Compton wavelength. Although in this regime the density of 
states vary with the virtuality $\lambda$, we will for simplicity ignore the
variations in our case~\cite{NOTE2}.  
For short times, the return probability is about 
constant but small, say $p(t)\sim {\bf p} +{\cal O}(t^2/\tau_d^2)$. Hence,
\be
R(s) = -\frac{{\bf p}}{\beta (\pi s)^2} 
\left(1-{\cal O} \left( \frac 1{(s\Delta\tau_d)^2}\right) \right).
\label{bal1}
\ee
A direct calculation of ${\bf p}$ may be averted by noticing that
(\ref{def}) implies
\be
\frac {d\,\Sigma_2 ( N)}{d N } = \int_{-N}^{N}
ds\, R(s)
\label{bal2}
\ee
for a fixed $N=E/\Delta$,
in the ballistic regime. Since (\ref{disc3}) obeys the
null sum rule on the {\em entire} support of eigenvalues as required by
the conservation of the number of energy levels, it follows that the 
contribution from the ballistic regime balances the contribution from the
diffusive regime for $L\gg l$, thereby fixing ${\bf p}$. Setting the cutoff
between the ballistic and diffusive regimes at $\sigma_l=D/(l^2\Delta)$, 
we obtain~\cite{NOTE3}
\be
R(s) =- \frac V{(2\pi l)^4 \beta}
\left( 1-{\bf C}+{\rm ln} (\frac {l^2}{D\tau_d})\right)
	\frac 1{s^2}
\label{bal4}
\ee
We note that $l^2/(D\tau_d)\sim 4$, and the prefactor yields ${\bf p}\approx 
V/l^4>0$, which is expected from free phase space in $d=4$.
In the virtuality range $\lambda \gg 1/\tau_d\sim m_Q$, the 
spectral rigidity for fixed $E/\Delta =N\gg 1 $, is
\be
\Sigma_2 ( N) = \frac {2V}{\beta (2\pi l)^4} \, 
\left(1-{\bf C} + {\rm ln} (\frac{l^2}{D\tau_d})\right)\,
{\rm ln}(N\Delta\tau_d)\,.
\label{bal5}
\ee
in overall agreement with the result of Gefen and Atland~\cite{GEFEN} for 
two-dimensional non-diffusive systems.

\vskip .25cm
{\bf 8.}
In models of the
QCD vacuum such as the instanton liquid model,  a chiral transition 
may be envisioned by varying the density of instantons in fixed $V$. 
The low density phase is characterized
by instanton clusters where the quarks and antiquarks are localized over the
cluster sizes. At high instanton density, 
the quarks are delocalized and chiral symmetry 
is spontaneously broken. These transitions are actually quite generic
and may be triggered by changing the external parameters in QCD,
including the shape and size of the Euclidean box $V$. For instance,
the formation of quark clusters is strongly flavor
dependent, so it is natural to assume that such a transition may take
place by just varying the number of flavors \cite{ZAKS,SHURYAKNF}. 
In the following, we will assume that these transitions are of the
metal-insulator type and proceed to analyze their signature on the
quark spectrum.

Let $\xi$ be the correlation or localization length in the cross-over
region. For $\xi\leq L$, the diffusion becomes anomalous with a scale
dependent diffusion constant~\cite{KRAV,ARONOV}. The scale dependence
maybe estimated using renormalization group arguments~\cite{LEE,STONE}.
At the critical point with $\xi\sim L$ and $s\gg 1$, the level correlation
asymptotes $R(s)\sim s^{-1-1/(\nu d)}$ \cite{KRAV,ARONOV}. The critical
exponent $\nu$ sets the rate of divergence of $\xi$ in terms of the critical 
conductance $\sigma_*$, $\xi/l\approx |1-\sg_l/\sg_*|^{-\nu}$~\cite{LEE}. 
In~\cite{KRAV} this result was reached by 
resumming the multi-2-diffuson diagrams and using renormalization group
analysis. Their arguments carry to QCD by trading 2-diffusons with 2-pions
(see above). In particular $d=4$, $\beta=2$, and $\sg_*\approx F_*^2\xi^2$.
From the $2+\epsilon$ expansion~\cite{LEE,STONE}, $\nu=1/2$
and the pion decay constant at the mobility edge $F_*$ vanishes with
universal exponent $\nu$.

At the critical point the normal
diffusive regime with constant $D$ disappears. Critical disorder is
characterized by $\sigma_*\sim \sigma_l$~\cite{LEE}. The spectral rigidity
in the regime : $\sigma_*< s<\sigma_* (\Delta_l/\Delta)$ with 
$\Delta_l=1/\rho\, l^2$, follows from the null spectral sum rule through
the ballistic contribution (\ref{bal4}). Hence,
\be
\Sigma_2 ( N) = 2 \frac {1-{\bf C}+{\rm ln 4}}{\beta (2\pi)^4}\,\frac 
{N}{\sigma_*} =\chi\, N\,\,.
\label{bal5n}
\ee
The level compressibility $\chi$
is weaker than Poisson ($\chi =1$) and universal.
This result is in agreement with the original analysis in~\cite{UNIVERSAL}
(see also~\cite{ARONOVMIRLIN}).

Finally, the asymptotics of $R(s)$ for $s\gg 1$ in the critical regime, implies 
that the level spacing distribution $P(s)$~\cite{MEHTA} in QCD changes 
at the mobility edge. The result is a modified Wigner surmise that is 
intermediate between Wigner-Dyson and Poisson~\cite{KRAV,ARONOV,UNIVERSAL}. 
This transition between chaos and integrability suggests that in the
same regime the quark wavefunctions in 
fixed $V$ are multi-fractal~\cite{FRACTAL} with a multi-fractal exponent 
$\eta=8\chi$.

\vskip .25cm
{\bf 9.}
The present results may be checked numerically by studying quark spectra
using lattice QCD simulations, or continuum 
models such as the the instanton liquid model or the Nambu-Jona-Lasinio 
model. They can be readily extended to matter since (\ref{dis1}) is protected
by symmetry, and supersymmetric gauge theories with flavor~\cite{SEIBERG}.

\vskip 0.5cm
IZ would like to thank Igor Aleiner for discussions.
This work was supported in part by the US DOE grant DE-FG-88ER40388, by the 
Polish Government Project (KBN) grants 2P03B04412 and 2P03B00814 and by the 
Hungarian grants FKFP-0126/1997 and OTKA-F026622.


\begin{thebibliography}{99}
\bibitem{NOTE1}
The general case with $\Lambda$ finite will be discussed elsewhere.
\bibitem{NOTE2}
The variations in the density of states can be taken into account by using
unfolding procedures~\cite{MEHTA}.
\bibitem{NOTE3}
The sensitivity of the result to the cutoff $\sigma_l$ is
logarithmically weak.
\end{thebibliography}
\end{document}